\documentstyle[epsf]{article}

\newcommand{\text}[1]{\rm #1}
\newcommand{\BE}{\begin{equation}}
\newcommand{\EE}{\end{equation}}
\newcommand{\BA}{\begin{eqnarray}}
\newcommand{\EA}{\end{eqnarray}}
\newcommand{\BAA}{\begin{eqnarray*}}
\newcommand{\EAA}{\end{eqnarray*}}

\begin{document}
\begin{titlepage}
\title{Wetting phenomena in bcc binary alloys$^1$}
\author
{
R.~Leidl,$^{2,3}$ A.~Drewitz,$^2$ and H.~W.~Diehl$^2$
}
\maketitle
\begin{abstract}
We study the influence of the surface orientation on the wetting behavior
of bcc binary alloys, using a semi-infinite lattice model
equivalent to a nearest-neighbor Ising antiferromagnet
in an external magnetic field. This model describes alloys that exhibit
a continuous $B2$--$A2$ order-disorder transition, such as FeAl or FeCo.
For {\em symmetry-breaking\/} surfaces like (100)
an {\em effective ordering surface field\/} $g_1\neq0$ emerges.
Such a field does not only crucially affect the surface
critical behavior at bulk criticality, but also gives rise
to wetting transitions below the critical temperature $T_c$.
Starting from the mean-field theory for
the lattice model and making a continuum approximation, a suitable
Ginzburg-Landau model is derived. Explicit results for the dependence
of its parameters (e.g., of $g_1$) on the microscopic interaction constants
are obtained. Utilizing these in conjunction with Landau theory,
the wetting phase diagram is calculated.
\bigskip

\end{abstract}
\vfill

\begin{picture}(5,1)
\unitlength1cm
\thicklines
\put(0,0){\line(1,0){4}}
\end{picture}

\begin{itemize}
\item[$^1$] Paper presented at the {\em Thirteenth Symposium on Thermophysical
Properties\/},\\ June 22-27, 1997, Boulder, Colorado, U.S.A.
\item[$^2$] Fachbereich Physik, Universit\"at-Gesamthochschule Essen,
D-45117 Essen,\\ Federal Republic of Germany.
\item[$^3$] To whom correspondence should be addressed.
\end{itemize}
\end{titlepage}

\newpage
\section{Introduction}\label{Introduction}
\indent

Surface critical behavior at bulk critical points
can be divided into distinct universality classes
\cite{Diehl_DL10}. For a given bulk universality class,
only gross surface properties determine which surface universality
class applies, such as: whether or not the surface interactions
exceed or are equal to a certain critical enhancement,
or whether a surface field $g_1$ coupling
to the local order parameter exists.
Recently it has been shown that the universal critical behavior
at the surface of a bcc Ising antiferromagnet
and of a binary alloy undergoing a continuous order-disorder bulk transition
depends crucially on the orientation
of the surface with respect to the crystal axes \cite{DrLDB,LD}.
The basic mechanism underlying this intriguing behavior
is the interplay between {\em broken translational invariance\/}
perpendicular to the surface and the symmetry with respect
to {\em sublattice ordering\/}. For certain ``symmetry-breaking''
orientations an ``effective'' ordering surface field $g_1\neq0$ emerges,
which depends on physical parameters like temperature and bulk composition
of the alloy.
That such a field exists has already been pointed out in \cite{S}
in order to explain the persistence of surface order at a (100) surface
above the bulk critical temperature $T_c$, detected in a Monte Carlo simulation for the $B2$--$A2$ order-disorder transition in Fe--Al.

The situation encountered for symmetry-breaking surfaces closely resembles
the critical adsorption of fluids, where generically $g_1\neq0$
\cite{crit_ads}. However, in that case the microscopic origin
of $g_1$ is quite different: It
is an external field reflecting, e.~g., the preference of the wall
for one of the two components of the binary liquid mixture.
The transition that takes place at the surface of the system
in the presence of a field $g_1\neq0$ on approaching the bulk critical point
has been called {\em normal\/} in \cite{BD}.
If $g_1=0$ (and the surface interactions are not too strongly enhanced),
another transition, called {\em ordinary\/}, occurs.
In accordance with
the fact that $g_1$ is a relevant scaling field,
the ordinary and normal transitions represent different surface universality
classes.

In Refs.\ \cite{DrLDB} and \cite{LD} the focus has been
on the behavior at $T=T_c$ and a clear identification
of the normal transition, which
may also be regarded as a critical point wetting phenomenon
\cite{Dietrich_DL12}. However, since $g_1$
{\em generally stays nonzero\/}
away from $T_c$ for symmetry-breaking surfaces,
a variety of wetting phenomena may occur for $T<T_c$.
Below we will determine the wetting phase diagram
for a (100) surface within the mean-field approximation,
utilizing the continuum model derived in \cite{LD}.
Our work complements previous studies
on wetting in fcc Ising antiferromagnets or binary alloys \cite{KG}
as well as on interface roughening at an antiphase boundary
in the [100] direction in bcc binary alloys \cite{SB}.

The organization of the paper is as follows. In the next section we
define our model, explain the difference between symmetry-breaking
and symmetry-preserv\-ing surfaces, and then briefly discuss
the discrete mean-field equations (Sec.\ \ref{Model}).
In Sec.\ \ref{Continuum} we introduce the Ginzburg-Landau model
for the (100) surface derived in \cite{LD}. This is then used
in Sec.\ \ref{Wetting} to determine the wetting phase diagram.

\section{Lattice model}\label{Model}

\subsection{Definition}
\indent

To model the continuous $B2$--$A2$ order-disorder
transition in the binary ($AB$) alloys FeAl or FeCo, we consider a bcc Ising antiferromagnet with nearest-neighbor (NN) interactions of strength
$J<0$ . The spin variable
$\sigma_i$ takes the values $+1$ or $-1$ depending on whether
lattice site $i$ is occupied by an $A$ or $B$ atom. 
Within the grand-canonical ensemble, the Ising Hamiltonian reads:
\BE\label{Hamiltonian}
{\cal H} = -J\sum_{\langle i,j\rangle}\sigma_i\sigma_j
- H\sum_i\sigma_i - H_1\sum_{i\in\text{surf}}\sigma_i\,,
\EE
where $\sum_{\langle i,j\rangle}$ runs over all  NN bonds.
The bulk field $H$ serves to adjust the composition of the alloy
and represents the chemical potential difference between $A$ and $B$ atoms.
A nonzero surface field $H_1$ occurs generically in binary alloys, giving rise
to surface segregation effects. One has $H_1>0$ ($H_1<0$) if $A$ ($B$) atoms
tend to segregate at the surface.
It is important not to confuse $H_1$ with an ``ordering''
field, which couples directly to the local order parameter.
Only if $H_1$ distinguishes one of the two sublattices at the surface,
as is the case for symmetry-breaking surface orientations where the surface sites
belong to a single sublattice only (see below), will $H_1$  contribute
to an ``effective'' ordering field $g_1$.
But even then $H_1\neq0$ is not a necessary condition
for $g_1\neq0$ (cf.\ Sec.\ \ref{Continuum}).

The average concentration (or occupation probability) $c_i$ of $A$ atoms
on lattice site $i$ can be written in terms of the mean magnetization
$m_i\equiv\langle\sigma_i\rangle$ of spin $\sigma_i$ as
$c_i=\frac{1}{2}(1+m_i)$. In the ordered ($B2$) phase,
the bcc lattice splits into two interpenetrating sc sublattices $\alpha$
and $\beta$ with bulk magnetizations $m_b^\alpha\neq m_b^\beta$,
which are preferentially occupied by $A$ and $B$ atoms, respectively
(cf.\ Fig.\ 1). The disordered ($A2$) structure is characterized
by $m_b^\alpha=m_b^\beta=:m_{\text{dis}}$.
The bulk order parameter is defined by
\BE\label{OPbulk}
\phi_b\equiv\frac{1}{2}\left(m_b^\alpha-m_b^\beta\right)\,.
\EE

\subsection{Symmetry properties of the surface}\label{Orientations}
\indent

Let us define more precisely what is meant by symmetry-breaking
and sym\-metry-preserving surfaces.
Consider a uniform translation $\tau_{\alpha\beta}$ of the crystal lattice
that maps $\alpha$-sites into $\beta$-sites.
In an infinite system without free surfaces or a finite system with
periodic boundary conditions, the Hamiltonian
${\cal H}={\cal H}\{\sigma_i\}$, Eq.\ (\ref{Hamiltonian}),
is {\em invariant} under the transformation:
\BE\label{symm_op}
\sigma_i\to\sigma_i^\prime=\sigma_{i+\tau_{\alpha\beta}}\,.
\EE
This symmetry is spontaneously broken below $T_c$
where the mean magnetizations $m_b^\alpha$ and $m_b^\beta$
transform into each other under (\ref{symm_op}),
so that $\phi_b\to-\phi_b$.
For a system with a free surface, the invariance of the Hamiltonian
still holds if $\tau_{\alpha\beta}$
can be taken {\em parallel} to the surface
as is the case for the (110) orientation (Fig.\ 1).
Then the surface is called {\em symmetry-preserving}.
One may convince oneself that a surface with Miller indices
$(n_1 n_2 n_3)$ is symmetry-preserving if and only if $n_1+n_2+n_3$ is even.
(By convention, we use the cubic unit cell of the bcc lattice here.)
The order parameter, which becomes a local quantity
$\phi=\phi_n$ depending on the discrete layer index $n$,
vanishes identically above $T_c$ since the $\phi\to-\phi$
symmetry of the bulk system is retained and no enhanced surface
interactions have been admitted in (\ref{Hamiltonian}).
By contrast, if either one of the lattice planes parallel to the surface
belongs to a single sublattice, as is the case for the (100)
surface (Fig.\ 1), {\em no} translation $\tau_{\alpha\beta}$
parallel to the surface exists
and the surface is called {\em symmetry-breaking}.
Generally, a $(n_1n_2n_3)$ surface is symmetry-breaking 
if $n_1+n_2+n_3$ is odd. Then the Hamiltonian
is no longer invariant under (\ref{symm_op}) but changes
by an amount proportional to the total number of surface sites.
(One may again consider a finite system but impose periodic boundary conditions
only in the directions parallel to the surface.)
Thus the $\phi\to-\phi$ symmetry of the bulk system is generically broken,
and the order parameter
will be nonvanishing at least locally near the surface even if $T\ge T_c$.

The symmetry properties of the surface must also show up
in the context of suitable continuum (Ginzburg-Landau) models.
Let  $\phi_s$ be the value of the order parameter at the surface.
For symmetry-preserving surfaces,
the surface contribution to the Landau free energy will
only contain {\em even\/} powers of $\phi_s$.
By contrast, arbitrary {\em odd\/}
powers are expected to occur for symmetry-breaking surfaces
due to the loss of the $\phi\to-\phi$ symmetry.
In particular, the coefficient  of the linear term
may be identified with an ``effective'' ordering surface field $g_1\neq0$.
Of course, in order to estimate
the magnitude of $g_1$ and its dependence on physical parameters
such as temperature and bulk composition,
the  parameters of the continuum model must be related explicitly
to  lattice quantities (see Sec.\ \ref{Continuum}).

\subsection{Mean-field (Bragg-Williams) approximation}
\indent

Owing to the spatial inhomogeneity along the $z$ axis perpendicular
to the surface, exact treatments
of the model (\ref{Hamiltonian}) are very hard, and one usually has to rely
on approximate techniques such as the mean-field (MF) or Bragg-Williams
approximation. The MF equations read (with $k_B =$ Boltzmann's constant):
\BE\label{MF}
m_i = \tanh\!\left[\frac{1}{k_BT}\left(H_i
-J\,\mbox{$\displaystyle\sum\limits_j$}^{(i)}m_j\right)\right]\,,
\EE
 where  $H_i=H$,
$H_i=H+H_1$ for  bulk and surface sites,
respectively. The sum $\sum^{(i)}$ runs over all NN sites of $i$.
The mean magnetizations $m_i$ vary with the index $n=1,2,\ldots$
labeling the lattice planes along the $z$ axis,
but are the same on each sublattice within a layer.
Thus for the (110) surface, {\em two\/} variables are needed to describe the state
of each layer:
\BE
m_i\equiv m_n^\alpha\quad\mbox{for $i\in$ layer $n$, subl.\ $\alpha$}\,,\quad
m_i\equiv m_n^\beta\quad\mbox{for $i\in$ layer $n$, subl.\ $\beta$}\,.
\EE
The local order parameter $\phi_n$ is conveniently defined as
\BE\label{OPloc}
\phi_n\equiv\frac{1}{2}\left(m_n^\alpha-m_n^\beta\right)\,.
\EE
For the (100) surface, one may write
\BE
m_i\equiv m_n\quad\mbox{for $i\in$ layer $n$}\,,
\EE
since each lattice plane belongs to a single sublattice.
The definition
of the local order parameter requires more care. The obvious choice
$\phi_n\equiv\frac{1}{2}(-1)^{n+1}(m_{n}-m_{n+1})$ is physically reasonable
but causes considerable problems in the continuum limit, as explained
in detail in \cite{LD}. It is favorable
to adopt the more symmetric definition \BE
\phi_n\equiv\frac{1}{2}(-1)^{n+1}
\left[m_n-\frac{1}{2}(m_{n+1}+m_{n-1})\right]\,,
\EE
which treats the preceding and succeeding layer
on an equal footing.

The MF equations  (\ref{MF}) have been studied in \cite{LD}
both for (110) and (100) surfaces via the ``nonlinear-mapping''
technique \cite{PW}. 
It has been shown that in the case of the (100) surface a nonvanishing order parameter profile appears for $T>T_c$. The characteristic length
scale  that governs its exponential decay at large $z$ 
{\em diverges\/} as $T\to  T_c$. Precisely at $T=T_c$,
the decay takes the form $z^{-\beta/\nu}$
(where $\beta=\nu=1/2$ within MF theory),
which is one of  the signatures of the normal transition.
In the case of the (110) surface, the local order parameter vanishes for $T>T_c$
since $m_n^\alpha=m_n^\beta$. Nevertheless one
obtains a nontrivial magnetization profile, so that
 $A$ rich and $B$ rich layers alternate as one moves along the $z$ direction.
However, the length scale associated with this profile remains {\em finite\/}
at $T=T_c$.

\section{Continuum (Ginzburg-Landau) model}\label{Continuum}
\indent

In proceeding to a suitable continuum description
one must be aware of several novel features arising for Ising antiferromagnets
or binary alloys, which are not present in simpler systems equivalent
to the Ising ferromagnet. First, 
the Landau expansion of the surface free energy should look
different for distinct orientations,
due to the loss of the $\phi\to-\phi$ symmetry for symmetry-breaking surfaces
(cf.\ Sec.\ \ref{Orientations}).
Second, one has to take into account so-called ``non-ordering''
(or noncritical) densities. In the alloy picture these are needed
to account for spatially varying profiles of, e.~g., the local
concentration that could {\em not\/} be described by the order parameter alone.

Non-ordering densities introduce {\em additional\/} length scales that may
{\em compete\/} with the order parameter correlation length $\xi_b$.
It has been shown in a study of wetting in fcc Ising
antiferromagnets that this competition may even lead to nonuniversal
critical wetting exponents,
and that the number of non-ordering densities depends on
the orientation of the surface. Of course,
noncritical densities do not affect the asymptotic surface critical behavior
if the bulk transition is continuous (as in our case of a bcc antiferromagnet),
since then the diverging correlation length $\xi_b$ dominates all other length
scales. However, we are interested here in phenomena
{\em below} $T_c$, where $\xi_b$ is finite.
Accordingly, length scales associated with such
spatially varying non-ordering densities may well be of the same order and
important.

To become more specific, let us recall the Ginzburg-Landau model
derived and critically examined in Ref.~\cite{LD} for the case
of the {\em symmetry-breaking\/} (100) surface. This
is based on a free-energy functional of the form
\BE\label{LG}
{\cal F}\{\phi\} =
\int_0^\infty dz\left\{\frac{c}{2}\left(\frac{d\phi}{dz}\right)^2
+f_b\left[\phi(z)\right]\right\} + f_s(\phi_s)\,,
\EE
where $\phi_s\equiv\phi(0)$.
The Landau expansions of the bulk and surface free-energy densities read
\BE\label{Landau_exp}
f_b(\phi) = \frac{a}{2}\phi^2 + \frac{b}{4}\phi^4
+ {\cal O}(\phi^6)\,,\quad
f_s(\phi_s) = -g_1\phi_s + \frac{c}{\lambda}\phi_s^2
+ {\cal O}(\phi_s^3)\,.
\EE
As expected, arbitrary odd powers of $\phi_s$ occur in the expansion
of $f_s(\phi_s)$. In view of the above discussion it is remarkable that
{\em no} spatially varying non\-ordering density appears in (\ref{LG}).
The reason is that lattice planes belonging to sublattice $\alpha$
and $\beta$ alternate along the [100] (or $z$) direction.
Hence the order parameter profile uniquely determines
the segregation profile and vice versa.
By contrast, the nontrivial segregation profile present
for the (110) surface above $T_c$ provides a typical example
of a spatially varying non-ordering density $\psi$
appearing in the corresponding free-energy functional \cite{L}.
The ``segregation field'' $H_1$ then couples linearly to $\psi$,
but no terms linear in the local order parameter and thus no
ordering surface field are present.

Deriving the continuum theory from the lattice model has the virtue
of relating the ``phenomenological'' coefficients
in (\ref{Landau_exp}) explicitly to the ``microscopic'' parameters:
\BE
K\equiv\frac{4|J|}{k_BT}\;,\;h\equiv\frac{H}{4|J|}\;,\;h_1\equiv\frac{H_1}{4|J|}\,,
\EE
which are dimensionless measures
of the spin coupling strength and the bulk and surface magnetic fields.
The ``bulk'' coefficients $a$, $b$, and $c$ are independent of surface
properties and depend on $K$ and $h$ only: $a=a(K,h)$ etc.
As usual, $a$ varies linearly with the reduced temperature
$t=(T-T_c)/T_c$ near $T_c$ (at fixed magnetic field $h$),
whereas $b$ and $c$ are positive constants to lowest order in $t$
(see \cite{LD}). The ``surface'' parameters are found to be:
\BE\label{g1}
\lambda(K,h)=1\,,\;g_1(K,h,h_1)=h_1+m_{\text{dis}}(K,h)\,,
\EE
where $m_{\text{dis}}=m_{\text{dis}}(K,h)$
is the magnetization of the disordered state. The latter
is thermodynamically stable only for $T>T_c$.
The surface field $h_1$ enters only in the ``effective''
ordering field $g_1$. The so-called extrapolation length
$\lambda$ (whose inverse $c_0\equiv1/\lambda$ is conveniently
called surface enhancement) is positive as it should be
if the surface interactions are not enhanced.

In order to better understand the expression for $g_1$ it is helpful
to recall some general symmetry properties of the Ising Hamiltonian
(\ref{Hamiltonian}) which should be respected by the continuum theory.
If one replaces $h$ and $h_1$ by its negative, the mean magnetizations
and the local order parameter (\ref{OPloc}) behave
as $m_i\to-m_i$, $\phi_n\to-\phi_n$, implying that
the ordering surface field should change sign, too:
\BE\label{g1_requ1}
g_1(K,-h,-h_1)=-g_1(K,h,h_1)\,.
\EE
For bulk field $h=0$ and arbitrary $h_1$,
the Ising antiferromagnet is {\em exactly\/}
equivalent to an Ising ferromagnet 
since flipping all Ising spins on one sublattice
and changing the sign of $K$ leaves the
partition function invariant. For the semi-infinite ferromagnet, however,
an ordering field $g_1$ is merely equivalent to a surface magnetic field
acting on the spins of the first layer and one easily finds \cite{PW}:
\BE\label{g1_requ2}
g_1(K,0,h_1) = h_1\,.
\EE
The expression (\ref{g1}) fulfills both (\ref{g1_requ1}) and (\ref{g1_requ2}).
That the bulk magnetization $m_{\text{dis}}$ of the disordered phase
comes into play can be understood as follows. For $T>T_c$,
the order parameter profile should vanish identically if $g_1(K,h,h_1)=0$.
Hence the layer magnetization profile $m_n$
must be a constant, $m_n\equiv m_{\text{dis}}(K,h)$.
By Eq.\ (\ref{MF}),
the molecular fields acting on surface and bulk spins are
$H+H_1-4|J|\,m_2$ and $H-4|J|(m_{n-1}-m_{n+1})$, respectively.
Thus the flat profile is a solution if and only if
$h_1=-m_{\text{dis}}(K,h)$. Below $T_c$ a thermodynamically unstable
solution $m_n\equiv m_{\text{dis}}$ (i.~e.\ $\phi_n\equiv0$)
still exists if and only if $h_1=-m_{\text{dis}}(K,h)$.
In order that this unstable solution survives the continuum approximation,
one again has to demand that $g_1(K,h,h_1)=0$ if $h_1=-m_{\text{dis}}(K,h)$.

\section{Wetting phase diagrams}\label{Wetting}
\indent

The physical picture behind the wetting behavior at a (100) surface
is the following. Below $T_c$, two ordered bulk phases $\pm\phi_b$ coexist.
If, e.~g., the effective ordering field $g_1$ is positive,
the surface favors the phase $\phi_b>0$, i.~e.\ $A$ atoms tend to occupy
sublattice $\alpha$ planes $n=1,3,\ldots$ while $B$ atoms reside
preferentially on the $\beta$ planes $n=2,4,\ldots$.
However, it may occur that deeper in the bulk the role
of the two sublattices is interchanged and the order parameter
assumes the value $-\phi_b$ there.
Then an antiphase boundary separating two regions of ordered phase
appears. Antiphase boundaries always
occur in real alloys below the ordering temperature. They
are the analogs of domain walls in an Ising ferromagnet.
Now one may ask how the thickness of this ``adsorbed'' layer
of bulk phase behaves when the temperature is varied.
If the interface stays within a finite distance from the surface
for $T<T_w$ while moving
arbitrarily away into the bulk for $T>T_w$, a wetting transition
takes place at $T_w=T_w(h_1,h)$. The thickness of the layer
may either grow continuously as $T\uparrow T_w$ (second-order wetting
transition), or jump from a finite value below $T_w$ to infinity for $T\ge T_w$
in a first-order transition.

We have calculated
the wetting phase diagram in the space of physical parameters
$1/K\propto T$, $h$, and $h_1$. To this end we employed the full expressions
for the bulk and surface Landau free energies $f_b$ and $f_s$ in (\ref{LG})
to be found in \cite{LD}, and used
the standard equal-area construction to locate the transitions \cite{Cahn}.
Of course, a description in terms of continuum mean-field theory
is only sensible above the roughening temperature $T_R$ \cite{PSW},
where $T_R\simeq\frac{1}{3}T_c$ for $h=1$ \cite{SB}.
For $T<T_R$ the growth of the wetting layer
proceeds via an infinite sequence of layering transitions
which are outside the scope of the continuum theory.
Fig.\ 2 shows two representative phase diagrams
at fixed bulk fields $h=1$ and $h=1.5$.
The tricritical point separating continuous and first-order transition lines
on the right branch of the phase diagram (where $g_1>0$)
is found to depend strongly on the bulk field
for $h$ larger than $\approx1.5$.
If $|h|>2$, the ordered phases are energetically unstable at $T=0$
and the bulk phase transition ceases to exist.
In order to interpret Fig.\ 2 an exact groundstate analysis
may be carried out similar to the one in \cite{PSW},
from which one easily finds that complete wetting already occurs at $T=0$
if $|h_1|\ge1$. Thus continuum mean-field theory is expected to fail
if wetting transitions at finite temperatures are predicted for $|h_1|>1$.
In this case one has to resort to the discrete mean-field theory
for a more accurate description of the low-temperature part
of the phase diagram.
\bigskip

\noindent
{\Large\bf Acknowledgments}

\noindent
This work has been supported by the Deutsche Forschungsgemeinschaft
via Sonderforschungsbereich 237 and the Leibniz program.

\newpage
\noindent
{\Large\bf Figure captions}
\bigskip

Fig.~1. Disordered ($A2$) and ordered ($B2$) structures
of a bcc binary ($AB$) alloy.
Black and white circles denote $A$ and $B$ atoms (or sublattice $\alpha$
and $\beta$ sites), respectively.
The hatched planes represent the (100) and (110) surface orientations.
\medskip

Fig.~2. Wetting phase diagrams at fixed bulk fields $h=1$ and $h=1.5$,
exhibiting continuous (full lines) and first-order
wetting transitions (dashed lines). One has $g_1>0$ ($g_1<0$)
to the right (left) of the dashed-dotted lines where $g_1=0$.

\newpage\thispagestyle{empty}
\hfill
\large\bf Figure 1 (Leidl/Drewitz/Diehl)
\vfill

\epsfxsize=\textwidth
\centerline{\epsffile{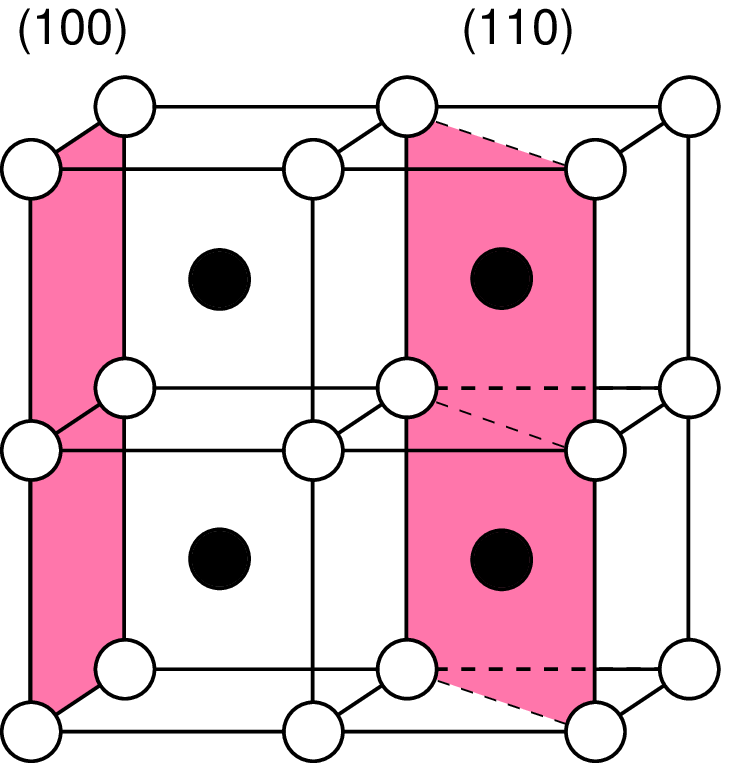}}

\newpage\thispagestyle{empty}
\hfill
\large\bf Figure 2 (Leidl/Drewitz/Diehl)
\vfill

\epsfxsize=0.9\textwidth
\centerline{\epsffile{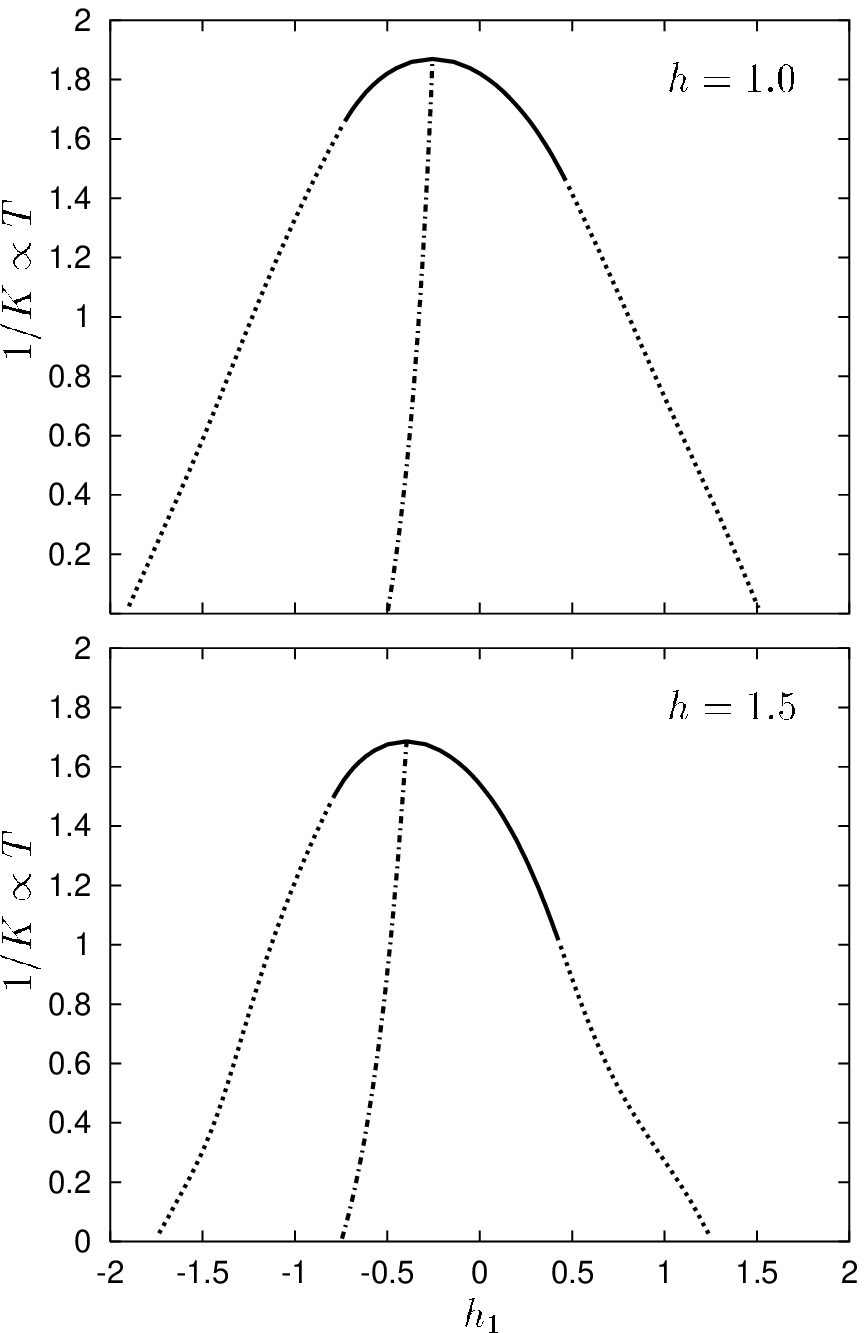}}


\begin{thebibliography}{99}
%
\bibitem{Diehl_DL10}{H.~W.\ Diehl, in {\it Phase Transitions and Critical
Phenomena},  edited by C.\ Domb and J.~L.\ Lebowitz (Academic, London, 1986),
Vol.\ 10, p.\ 75.}
%
\bibitem{DrLDB}{A.\ Drewitz, R.\ Leidl, T.~W.\ Burkhardt, and H.~W.\ Diehl,
{\it Phys.\ Rev.\ Lett.}\ {\bf 78}:1090 (1997).}
%
\bibitem{LD}{R.\ Leidl and H.~W.\ Diehl, preprint.}
%
\bibitem{S}{F.\ Schmid, {\it Z.\ Phys.\ B} {\bf 91}:77 (1993).}
%
\bibitem{crit_ads}{H.~W.\ Diehl,
{\it Ber.\ Bunsenges.\ Phys.\ Chem.}\ {\bf 98}:466 (1994).}
%
\bibitem{BD}{T.~W.\ Burkhardt and H.~W.\ Diehl,
{\it Phys.\ Rev.\ B} {\bf 50}:3894 (1994).}
%
\bibitem{Dietrich_DL12}{S.\ Dietrich, in {\it Phase Transitions and Critical
Phenomena},  edited by C.\ Domb and J.~L.\ Lebowitz (Academic, London, 1988),
Vol.\ 12, p.\ 1.}
%
\bibitem{KG}{D.~M.\ Kroll and G.\ Gompper,
{\it Phys.\ Rev.\ B} {\bf 36}:7078 (1987).
G.\ Gompper and D.~M.\ Kroll, {\it Phys.\ Rev.\ B} {\bf 38}:459 (1988).}
%
\bibitem{SB}{F.\ Schmid and K.\ Binder,
{\it Phys.\ Rev.\ B} {\bf 46}:13553 (1992); {\it ibid.}, 13565 (1992). }
%
\bibitem{PW}{R.\ Pandit and M.\ Wortis,
{\it Phys.\ Rev.\ B} {\bf 25}:3226 (1982).}
%
\bibitem{L}{R.\ Leidl, to be published.}
%
\bibitem{Cahn}{J.~W.\ Cahn, {\it J.\ Chem.\ Phys.}\ {\bf 66}:3667 (1977).}
%
\bibitem{PSW}{R.\ Pandit, M.\ Schick, and M.\ Wortis,
{\it Phys.\ Rev.\ B} {\bf 26}:5112 (1982).}
%
\end{thebibliography}
\end{document}